# BLOCKCHAIN BASED DIGITAL VACCINE PASSPORT

Ms. MEGHA RANI R
Asst.Professor
Dept. Of CSE
VTU
Megharani.cs@sahyadri.edu.in

ROSHAN R ACHARYA
Dept. Of CSE
VTU
Roshanrajnishacharya1@gmail.com

RAMKISHAN
Dept. Of CSE
VTU
kishan.ram1221@gmail.com

RANJITH K
Dept. Of CSE
VTU
Ranjithk2610@gmail.com

RAKSHITH AY GOWDA
Dept. Of CSE
VTU
rakshith1605@gmail.com

*Abstract—* **Travel has been challenging recently since different nations have implemented varied immigration and travel policies. For the time being, immigration officials want proof of each person's immunity to the virus. A vaccine passport serves as evidence that a person has tested negative for or is immune to a particular virus. In terms of COVID-19, those who hold a vaccine passport will be permitted entry into other nations as long as they can provide proof that they have COVID-19 antibodies from prior infections or from full COVID-19 immunizations. To reduce time and effort spent managing data, the vaccination passport system has been digitalized. The process of contact tracing may be facilitated by digitization. The "Blockchain technology" system, which is currently in use, has demonstrated its security and privacy in systems for data exchange among bitcoin users. The Digital Vaccination Passport scheme can use Blockchain technology. The end result would be a decentralized, traceable, transparent, reliable, auditable, secure, and trustworthy solution based on the Ethereum block-chain that would allow tracking of vaccines given and the history of diseases**

I. INTRODUCTION

A COVID-19 vaccine is one that aims to give acquired immunity against the coronavirus disease-causing virus known as severe acute respiratory syndrome coronavirus 2 (SARS CoV 2). (COVID-19). The initial COVID-19 vaccinations were created and made accessible to the general population in 2020 thanks to emergency use authorization. The COVID-19 vaccines are widely credited with helping to stop the disease's spread as well as its severity and fatality rates. Many nations adopted staggered distribution strategies that gave priority to individuals who were most vulnerable to problems, such the elderly, and to exposure and transmission risks, like healthcare personnel. According to official statistics from national public health organisations, 11.81 billion doses of the COVID-19 vaccination have been given worldwide as of May 26, 2022. By December 2020, more than 10 billion vaccine doses had been pre-ordered by nations, with 14 percent of the world's population, or roughly half of the doses, coming from high-income nations.

The part of vaccination policy that pertains to persons crossing borders is vaccination requirements for international travel. In order to stop epidemics, several nations throughout the world demand that visitors who are going to or coming from other nations have the necessary vaccinations. These travellers must present confirmation of vaccination against particular diseases at border crossings.

An immunity passport known as a vaccine passport or proof of vaccination is used as identification in some nations and jurisdictions as part of the vaccination-based COVID-19 pandemic control strategy. A government or health body normally issues a vaccine passport, which might be a digital or physical document. Some credentials might have a QR code that can be read by a smartphone app for provisioning. The idea behind vaccination passports is that those who have received the vaccine are less likely to spread the SARS-CoV-2 virus to others and are less likely to suffer a serious consequence (hospitalisation or death) if they become ill, making it safer for them to congregate. A vaccine passport is generally coordinated with corporate policies that are implemented or enforceable public health orders that demand customers provide proof of COVID-19 vaccination as a requirement of access or service.

Many governments are thinking about developing contemporary, digital vaccine passports that are more difficult to fake, even though vaccination cards like yellow cards are still in use and are still a common means to track immunizations. Multiple nations are investigating whether vaccine passports and health permits could be used as proof of COVID-19 vaccination in light of the public health hazard posed by the COVID-19 pandemic. This would let individuals resume their regular activities and restore confidence in international travel. With new, secure digital credentialing technologies becoming more common, vaccine passports are benefiting from this development. The use of it extends beyond vaccine passports for international travel to various contexts. Organizations that collect individuals in groups, for instance, are looking for digital alternatives to paper test results and immunisation records. In some circumstances, this entails determining if people have had voluntary, privacy-preserving testing or vaccinations. A practical and optional way for people to report their health status, such as whether they have had a vaccination or have tested negative for COVID-19, is through a digital vaccine passport. People with digital health passes can share a scannable QR code on their smartphone that confirms their status without having to remember to carry around multiple documents, and personal information is kept securely encrypted in a digital wallet on the user's phone. Digital vaccine passports are one of many methods that governments, commercial firms, non-profits, and industry groups are proposing to enable individuals get back to their favourite hobbies as COVID-19 vaccine rollouts get underway around the world.

Decentralized and transaction-based data exchange across a large network of participants is made possible by blockchain technology. For transactions that include unreliable parties and require high security, blockchain technology has recently attracted more academic focus. Data that has been entered into the blockchain cannot be changed once it has been done so because to the integration of the digital vaccination passport with blockchain technology. The Blockchain technology can play a crucial part in the relief of sickness as well as the facilitation of the implementation of governmental regulations and standards, maintaining trust between all stakeholders, in order to assist combat this global health crisis. In fact, the emerging Blockchain technology, which is a global computational infrastructure (i.e., smart contracts) and a distributed, immutable, and tamper-proof ledger database, has the potential to offer effective COVID-19 solutions based on high standards of accuracy and trust thanks to its essential characteristics of transparency, integrity, and resilience. In light of this, we suggest in this paper a Blockchain-based platform for issuing and confirming COVID-19 test/vaccine certificates called Digital Vaccine Passport.

II. IMPLEMENTATION

We have implemented the proposed framework using the following technologies:

1. ETHEREUM

A decentralised, open-source blockchain with smart contract capabilities is called Ethereum. The site uses Ether (ETH) as its native coin. Programmer Vitalik Buterin created Ethereum in 2013. Ethereum 2.0, often known as Eth2, is currently[may be out of date as of March 2022] under open-source development. The primary goal of the update is to boost the network's transaction throughput from its present pace of roughly 15 transactions per second[citation needed] to perhaps as much as tens of thousands of transactions per second.

Ethereum is a permissionless, non-hierarchical network of computers (nodes) that constructs and reaches consensus on the blockchain, a continually expanding collection of "blocks" or groups of transactions. Each block has a unique identifier for the chain that must come before it in order for it to be regarded as genuine. The ETH balances and other storage values of Ethereum accounts are changed whenever a node adds a block to its chain by carrying out the transactions in the block in the order they are specified. The "state," or collection of these balances and values, is kept on the node independently of the blockchain in a Merkle tree.

2. SOLIDITY

Smart contracts can be implemented using Solidity, a high-level programming language that is contract-oriented. C++, Python, and JavaScript have heavily inspired the development of Solidity, which was made with the Ethereum Virtual Machine in mind (EVM).

The runtime environment for smart contracts in Ethereum is the Ethereum Virtual Machine, or EVM. The Ethereum Virtual Machine is focused on ensuring security and enabling machines all across the world to run untrusted code.

The EVM is specialised in thwarting denial-of-service attacks and guarantees that applications cannot access each other's state, allowing for uninterrupted communication. The Ethereum Virtual Machine was created to operate as a runtime environment for Ethereum-based smart contracts.

3. FIREBASE

User authentication is one of the most crucial criteria for Android apps in the modern era. User authentication is crucial, but if we have to create all of this code by hand, it will be much more difficult.

With Firebase's assistance, this is accomplished relatively quickly. The FirebaseUI allows us to sign people into our app. It manages the UI processes for logging in using an email address and password, phone numbers, and widely used providers, such as Google Sign-In and Facebook Login. Also possible are situations like account recovery. UI design is not necessary because one is already available to us. It implies that the activities need not be recorded.

Before we can login someone into our app, we need to collect their authentication credentials. The user's email address and password can serve as their credentials. An OAuth token from an identity provider can be used as the credential. We then provide the Firebase Authentication SDK with these credentials. After confirming such credentials, backend services will reply to the client.

4. FRONT END FRAMEWORKS

- HTML

Standard markup for documents intended to be seen in a web browser is called HTML, or HyperText Markup Language. Cascading Style Sheets (CSS) and JavaScript are examples of technologies that can help with this.

- CSS

A style sheet language called Cascading Style Sheets (CSS) is used to describe how a document produced in a markup language like HTML or XML is presented (including XML dialects such as SVG, MathML or XHTML). The World Wide Web's foundational technologies, along with HTML and JavaScript, include CSS.

- JAVASCRIPT

Along with HTML and CSS, the computer language known as JavaScript, or JS, is one of the foundational elements of the World Wide Web. By 2022, 98% of websites will employ JavaScript on the client side to control how web pages behave, frequently using third-party libraries. A dedicated JavaScript engine is available in every major web browser and is used to run the code on users' devices.

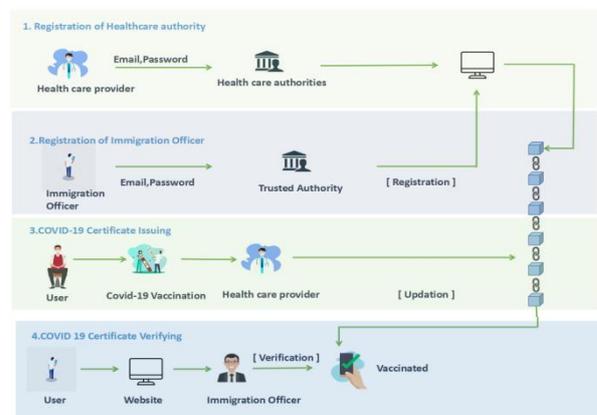

III. METHODOLOGY

There are four important phases in the operation of the proposed framework

1) Registration of healthcare authorities

2) Registration of immigration officer

3) Covid-19 certificate issuing

4) Covid-19 certificate verification

1. Registration of healthcare authorities:

In this phase to register himself, the healthcare provider provides his Email(registered with the hospital) and password to the concerned healthcare authorities. The healthcare authority will register the healthcare provider using the information provided, to the network through the UI provided. Now the healthcare provider can log in to the blockchain network through the login credentials provided at the time of registration

2. Registration of immigration officers:

Immigration officers are the registered personnel who are responsible for the verification of required documents of the traveller who is travelling from one place to the other. The immigration officer has to be registered to the network in order to do the aforementioned. The trusted authority collects the credentials provided by the immigration officer in order to register them to the network.

3. Covid-19 certificate issuing:

When the traveller get vaccinated at the hospital/healthcare front the healthcare provider will update the vaccination information of the traveller in the blockchain through the UI provided by the proposed framework.

The healthcare provider will be updating the following information onto the blockchain network:

1. Aadhar card number

2. Name as in registered ID

3. Vaccination information

4. Hospital name

1. Covid-19 certificate verification.

The immigration officer verifies the vaccination information of the traveller during the immigration process. The immigration officer does so by using the aadhar card number provided by the traveller. The immigration officer feeds the aadhar card number to the UI provided and searches for the vaccination record associated with the aadhar card number. Now the immigration officer can cross-verify the vaccination information of the traveller.

IV. RESULT AND DISCUSSION

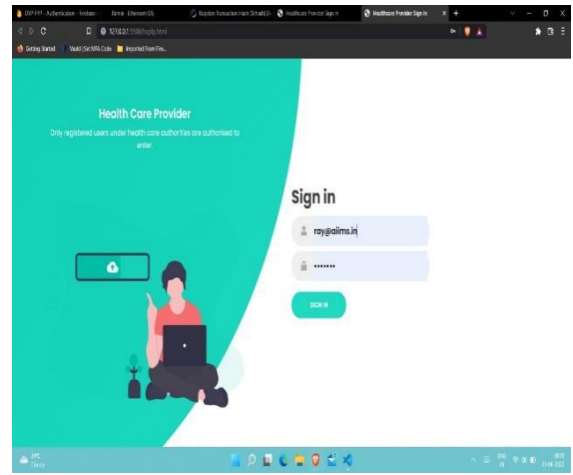

Figure 2: Authentication page

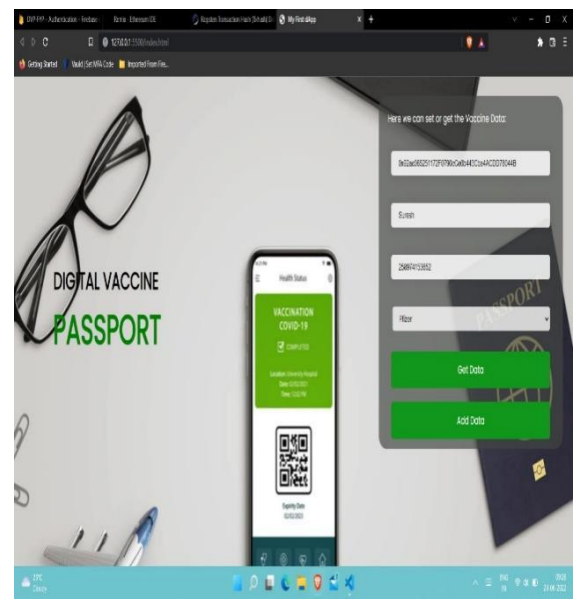

Figure 3: Landing page for data entry

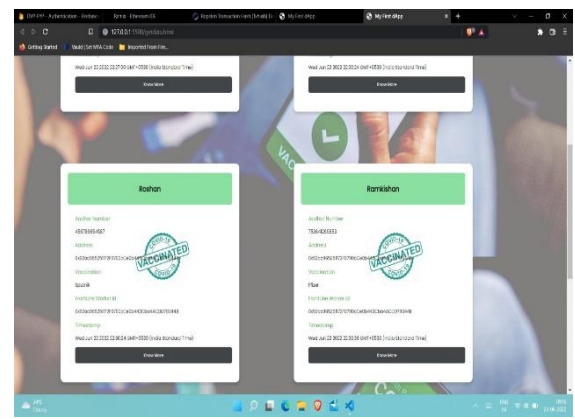

Figure 4: Landing page for immigration officers

## V. CONCLUSION AND FUTURE WORKS

Covid 19 certificates validates whether a particular individual is vaccinated or not. Travel constraints arises when covid 19 certificate is not submitted. These Certificates are made mandatory in the field like film theatres, hospitals, airports etc. which makes this a necessary document to carry with. So there are chances that these documents are tampered, or fake documents which is a challenge to the security of personal information. So, this literature focus on the concept of digital vaccine passport with the help of blockchain technology to generate the covid 19 vaccination certificates which are tamper-proof and cannot be faked. This takes user privacy in to account as well as security of documents. Since block chain is used with the use of smart contracts we can generate new secure block of data when new information is added. Proposed system allows the healthcare officials to register in the website and add the details of persons who are vaccinated and generate the certificate. Also, user can access the certificate using his credentials. This proposed system provides good security and privacy to user data by using blockchain technology

## VI. REFERENCES

[1] Dakota Gruener, 2020 Immunity Certificates: If We Must Have Them, We Must Do It Right

[2] Hye-Young Paik, Xiwei xu, H. M. N. Dilum Bandara, Sung Une Lee, Sin Kuang Lo, 2019 Analysis of Data Management in Blockchain-Based Systems

[3] Vitalik Buterin, 2017 Ethereum White Paper

[4] Ahmed Afif Monrat, Olov Schelén, Karl Andersson, 2019 A Survey of Blockchain From the Perspectives of Applications, Challenges, and Opportunities

[5] Mahdi H. Miraz, Maaruf Ali, 2018 Applications of Blockchain Technology beyond Cryptocurrency

[6] Fredrick Ishengoma, 2021 FC-Blockchain Based COVID-19 Immunity Certificate: Proposed System and Emerging Issues

[7] K. Christodoulou, P. Christodoulou, Z. Zinonos, E. G. Carayannis and S. A. Chatzichristofis, "Health Information Exchange with Blockchain amid Covid-19-like Pandemics," 2020 16th International Conference on Distributed Computing in Sensor Systems (DCOSS), 2020, pp. 412-417, doi: 10.1109/DCOSS49796.2020.00071.

[8] Abid A, Cheikhrouhou S, Kallel S, Jmaiel M. NovidChain: Blockchain-based privacy-preserving platform for COVID-19 test/vaccine certificates. Softw Pract Exper. 2021

[9] Fauzi Budi Satria, Mohamed Khalifa, Mihajlo Rabrenovic, Usman Iqbal, Computer Methods and Programs in Biomedicine Update

[10] José L. Hernández-Ramos, Georgios Karopoulos, Dimitris Geneiatakis, Tania Martin, Georgios Kambourakis, and Igor Nai Fovino, Sharing pandemic vaccination certificates through blockchain: case study and performance evaluation

[11] Mohammed Shuaib, Shadab Alam, Mohammad Shahnawaz Nasir, Mohammad Shabbir Alam, Immunity credentials using self-sovereign identity for combatingCOVID-19 pandemic

[12] Georgios Karopoulos, Jose L. Hernandez-Ramos, Vasileios Kouliaridis, Georgios Kambourakis, A Survey on Digital Certificates Approachesfor the COVID-19 Pandemic

[13] Don't Be Guilty of ThesePreventable Errors in Vaccine Administration!